\documentclass[aps,pra,twocolumn,superscriptaddress]{revtex4}

\usepackage{graphicx}
\usepackage{bm}
\usepackage{ulem}
\DeclareTextSymbol{\degre}{T1}{6}
\DeclareTextSymbol{\degre}{OT1}{23}
\DeclareGraphicsExtensions{.jpg,.eps}
\begin{document}

\newcommand{\lacuvo}{La$_3$Cu$_2$VO$_9$ }

\title{Formation of collective spins in frustrated clusters}
\author{J. Robert}
\affiliation{Institut NEEL, CNRS \& Universit\'e Joseph Fourier,
BP 166, F-38042 Grenoble Cedex 9, France}
\author{V. Simonet}
\affiliation{Institut NEEL, CNRS \& Universit\'e Joseph Fourier,
BP 166, F-38042 Grenoble Cedex 9, France}
\author{B. Canals}
\affiliation{Institut NEEL, CNRS \& Universit\'e Joseph Fourier,
BP 166, F-38042 Grenoble Cedex 9, France}
\author{R. Ballou}
\affiliation{Institut NEEL, CNRS \& Universit\'e Joseph Fourier,
BP 166, F-38042 Grenoble Cedex 9, France}
\author{E. Lhotel}
\affiliation{Institut NEEL, CNRS \& Universit\'e Joseph Fourier,
BP 166, F-38042 Grenoble Cedex 9, France}
\author{C. Darie}
\affiliation{Institut NEEL, CNRS \& Universit\'e Joseph Fourier,
BP 166, F-38042 Grenoble Cedex 9, France}
\author{P. Bordet}
\affiliation{Institut NEEL, CNRS \& Universit\'e Joseph Fourier,
BP 166, F-38042 Grenoble Cedex 9, France}
\author{B. Ouladdiaf}
\affiliation{Institut Laue-Langevin, BP 154, F-38042 Grenoble Cedex,
France.}
\author{M. Johnson}
\affiliation{Institut Laue-Langevin, BP 154, F-38042 Grenoble
Cedex, France.}
\author{J. Ollivier}
\affiliation{Institut Laue-Langevin, BP 154, F-38042 Grenoble
Cedex, France.}
\author{D. Braithwaite}
\affiliation{CEA-Grenoble, DRFMC/SPSMS/IMAPEC, 17 rue des Martyrs,
F-38054 Grenoble cedex, France}
\author{H. Rakoto}
\affiliation{Laboratoire National des Champs Magn\'etiques
Puls\'es, 143 avenue de Rangueil, F-31400 Toulouse, France}
\author{S. de Brion}
\affiliation{Grenoble High Magnetic Field Laboratory, CNRS, BP
166, F-38042 Grenoble, France}

\date{\today}

\begin{abstract}
Using magnetization, specific heat and neutron scattering
measurements, as well as exact calculations on realistic models,
the magnetic properties of the \lacuvo compound are characterized
on a wide temperature range. At high temperature, this oxide is
well described by strongly correlated atomic $S$=1/2 spins while
decreasing the temperature it switches to a set of weakly
interacting and randomly distributed entangled pseudo spins
$\tilde S=1/2$ and $\tilde S=0$. These pseudo-spins are built over
frustrated clusters, similar to the kagom\'e building block, at
the vertices of a triangular superlattice, the geometrical
frustration intervening then at different scales.
\end{abstract}

\maketitle

\section{Introduction}

A lot of work has been devoted recently to the understanding of
the peculiar magnetic behaviour of extended networks of spins on
triangle or tetrahedron based lattices. These elemental
configurations of spins, especially when they are corner-sharing,
imply strong geometrical frustration for antiferromagnetically
coupled Heisenberg spins, i.e. inability to simultaneously
minimize the pair interactions. The case of $S$=1/2 spins is
especially appealing because it should enhance quantum physical
behaviour. The archetypical frustrated network of spins in 2
dimensions is the kagom\'e lattice, which is expected to
stabilize, at low temperature, a spin liquid state \cite{kagome}.
One of the most striking feature of this state is the presence of
many low lying singlet states \cite{lecheminant}. This feature
could be interpreted in a short-range resonating valence bond
picture of the kagom\'e ground state \cite{mila}. Few experimental
realizations of kagom\'e lattice with $S$=1/2 spins are available.
Among these, the recently studied hebertsmithite \cite{hebert1}
does not present any magnetic transition down to the lowest
temperature but its fluctuating ground state is still under debate
since it does not meet the theoretical predictions
\cite{hebert1,hebert2,hebert3,lecheminant}. The theoretical
studies have been undertaken on ideal systems but exact solutions
can only be obtained for a limited number of spins (up to 36)
\cite{lecheminant,misguich}. This is one of the reasons why it
should be interesting to study experimentally a system constituted
of isolated frustrated clusters of spins, for which the exact
calculations are fully relevant.

The physics of spin clusters has attracted much attention by
itself. Quantum dynamics of mesoscopic magnets and decoherence
effects by specific environments for instance can be investigated
in organo-metallic molecular magnets with metal clusters
stabilizing a collective spin magnetically isolated by the organic
ligands. A number of investigations thus followed the discovery of
a staircase magnetization hysteresis loop associated with the
quantum tunneling of collective spins trough a magnetic anisotropy
energy barrier \cite{Mn12QT_1,Mn12QT_2}. Clusters with small total
spin and large N\'eel vector also raised strong interest. This is
the case of the polyoxovanadate, shortnamed V$_{15}$, constituted
of 15 V$^{4+}$ ions. Its magnetic properties, showing quantum
processes under sweeping field \cite{V15_1}, are accounted for by
the formation of a collective pseudo-spin $\tilde S$=1/2 from the
three weakly coupled $S$=1/2 spins on the central triangle of the
cluster \cite{V15_2}. The study of spin clusters within oxide
compounds was undertaken more recently. At variance with molecular
compounds, the magnetic screening then is much less effective,
which leads to non-negligible inter-cluster interactions and
therefore allows the study of the coupling of the collective
entities. For instance, in the oxide compound
La$_4$Cu$_3$MoO$_{12}$, a trimer of $S$=1/2 spins is stabilized in
each triangle of Cu$^{2+}$ at the vertices of a square lattice
\cite{LCMoO_1,LCMoO_2,LCMoO_3,LCMoO_4}. The inter-cluster
couplings in La$_4$Cu$_3$MoO$_{12}$ is two orders of magnitude
smaller than the intra-cluster ones, and lead to the onset of an
antiferromagnetic long-range order at 2.6 K. In both V$_{15}$ and
La$_4$Cu$_3$MoO$_{12}$ compounds, the splitting of the 2
fundamental $S$=1/2 spins doublets might be attributed to a
departure from the trigonal symmetry. A last remarkable oxide
compound whose properties are dominated by spin clusters is
Na$_2$V$_3$O$_7$ \cite{Na2V3O7_1,Na2V3O7_2,Na2V3O7_3,Na2V3O7_4}.
The clusters are built on rings of 9 antiferromagnetically
interacting V$^{4+}$ ions of $S$=1/2 spins, piled up in tubes with
inter-ring ferromagnetic interactions. One $S$=1/2 spin out of 9
is left active in the low temperature paramagnetic regime through
a dimerization process of the other spins of the ring,
characterized by a broad range of V-V interactions. The proximity
of a quantum critical point at $H$=0 is suggested from the complex
magnetic behaviour observed with respect to the magnetic field H
at low temperature T.

The layered oxide La$_3$Cu$_2$VO$_9$ \cite{jansson,vander}, parent
of the La$_4$Cu$_3$MoO$_{12}$ compound, provides a model at the
intersection of geometrical frustration and spin cluster research
areas. In the magnetic layers, the Cu$^{2+}$ ions form planar
clusters of antiferromagnetically coupled 8 and 9 $S$=1/2 spins,
arranged on 4 corner-sharing triangles. The geometrically
frustrated 9-spin cluster is very similar to the building block of
a kagom\'e network. As will be shown in the following, a
collective state of spin is constructed from the $S$=1/2 spins
wave functions of each cluster while decreasing the temperature.
The resulting collective pseudo-spins involve a greater number of
individual spins than in the V$_{15}$ or La$_4$Cu$_3$MoO$_{12}$
compounds. Like in La$_4$Cu$_3$MoO$_{12}$, the inter-cluster
couplings in La$_3$Cu$_2$VO$_9$ are two orders of magnitude
smaller than the intra-cluster ones, and their influence is
traceable in the experimental low temperature range. Moreover, the
clusters themselves are arranged on a triangular lattice, which
gives the opportunity to study the influence of geometrical
frustration at different spatial and energy scales in the same
compound.

In this article, following a preliminary report \cite{HFM}, we
will focus on the individual behavior of these pseudo-spins, their
interplay when inter-cluster interactions become effective will be
the subject of a forthcoming paper \cite{avenir}. The synthesis
and crystallography of \lacuvo polycrystalline sample will first
be described, with special emphasis on the Cu/V substitution.
Then, the magnetic measurements will be presented, in a wide
temperature range, following the construction of the collective
pseudo-spin, and also in high magnetic fields. The next two
sections will respectively present the results of specific heat
measurements and of inelastic neutron scattering. The last section
before concluding will be devoted to a discussion of these
experimental results in the light of exact calculations performed
on realistic models of spins cluster.

\section{\label{Exp}Experimental}

\subsection{\label{Sample} La$_3$Cu$_2$VO$_9$ sample characterization}

\subsubsection{\label{sec:synthesis} Synthesis and structure}

A polycrystalline \lacuvo sample was synthesized by a sol-gel
method. The stoichiometric metallic cations were dissolved in
nitric acid before being complexed by addition of EDTA in a
controlled pH solution. This solution was polymerized and then
heated at 700~$^{\circ~}$C to eliminate the organic constituents.
The resulting powder was annealed during 15 days at
1010~$^{\circ~}$C. The sample was first checked to be a single
phase using X-ray diffraction.

\begin{figure}
\includegraphics[scale=0.55]{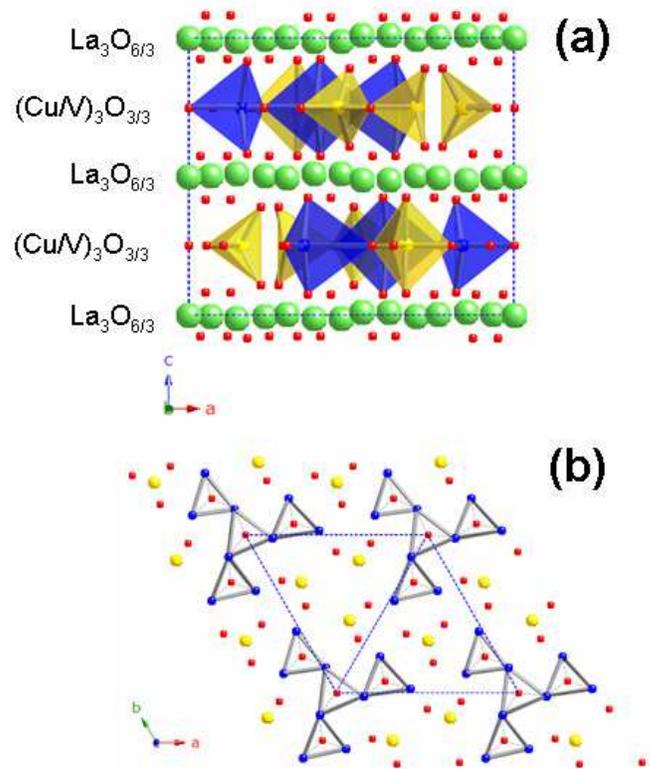}
\caption{(Color online) (a) La$_3$Cu$_2$VO$_9$ structure projected
along the b axis: LaO$_{6/3}$ layers alternating with
(Cu/V)O$_{3/3}$ ones with the O polyhedron drawn around Cu and V
atoms. (b) La$_3$Cu$_2$VO$_9$ structure projected along the c
axis: (Cu/V)O$_{3/3}$ layers showing the 9 Cu$^{2+}$ clusters at
the vertices of a triangular lattice. La, V, Cu and O atoms are
represented by large green, yellow, blue and small red circles,
respectively.}\label{fig:struct}
\end{figure}

\begin{figure}
\includegraphics[bb=20 400 572 714,scale=0.48]{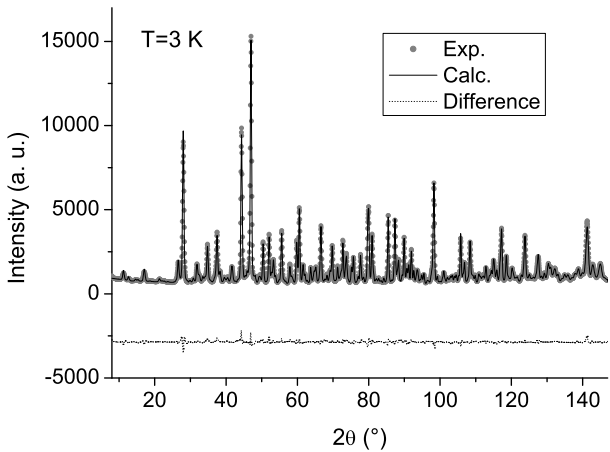}
\caption{\lacuvo powder neutron diffractogramm recorded at 3 K on
D2B, the corresponding refined pattern and their difference. The
refined structural parameters are presented in table
\ref{table1}.}\label{fig:neutrons}
\end{figure}

Neutron powder diffractograms were then recorded on the high
resolution powder diffractometer D2B at the Institut Laue Langevin
with a wavelength of 1.59 \AA\ using a Ge monochromator (cf. Fig.
\ref{fig:neutrons}). The absence of structural phase transition
was checked from 3 to 300 K. The crystallographic description of
the sample, obtained from the Rietvelt analysis of the neutron
diffractograms with the Fullprof program \cite{fullprof}, was
found consistent with the results of Ref. \cite{vander}: the
\lacuvo oxide compound crystallizes in the hexagonal P6$_3/m$
space group with lattice parameters a=b=14.395(2) \AA, c=10.656(2)
\AA\ at 300 K. As shown in Fig. \ref{fig:struct} (a), LaO$_{6/3}$
layers alternate with (Cu/V)O$_{3/3}$ layers underlying the 2D
structure of the compound. In these, the Cu$^{2+}$ ions, the only
magnetic ones with a $S$=1/2 spin, are localized on three
inequivalent sites Cu(2), Cu(3) and Cu(4) (following the site
labeling of ref. \onlinecite{vander}) having different
coordination environments. These are forming 9-spin planar
clusters, constituted with 4 corner-sharing triangles. The
clusters are centered at the vertices of a 2D triangular lattice
(cf. Fig. \ref{fig:struct} (b)). Within the planes, O-coordinated
V atoms are intercalated between the clusters.

A Cu/V substitution, of vanadium V$^{5+}$ ions for copper
Cu$^{2+}$ ions, dictated by the global charge neutrality
requirement that fixes the stoichiometric proportions of the
different ions, occurs in the very small ratio of $\frac{1}{27}$.
Notice nevertheless that this in average corresponds to one Cu/V
substitution over $\frac{1}{3}$ of the 9-spin planar clusters. We
will detail the nature and consequence of this substitution below.

In non substituted clusters, the neighbor Cu-Cu distances within
the cluster range between 3.35 \AA\ and 3.8 \AA, whereas the
shortest in-plane distances between the peripheral Cu of the
neighboring clusters is larger and equal to 4.55 \AA. In
comparison, the distance between two adjacent (Cu/V)O$_{3/3}$
layers is 5.3 \AA.

Within the (Cu/V)O$_{3/3}$ layers, the first neighbor $S$=1/2
spins of each cluster are interacting through superexchange via an
intermediate O. Concerning the inter-cluster magnetic interaction
within one layer or between adjacent layers, possible longer
superexchange paths can be effective through one or two O.

\subsubsection{\label{sec:substitutions}Cu$/$V Substitution}

In order to understand the magnetic properties of the compound, it
was found essential to take into account the Cu/V substitution and
the related modifications of the cluster structure.

To get insight into the distribution of the V in the clusters,
refinement of the powder neutron diffraction data at 3 K was made
allowing V substitution on each copper sites, alternatively.
Although the fit is slightly better when the Cu(2) site (central
triangle) is substituted instead of Cu(3) and Cu(4) sites, in
agreement with previous results \cite{vander} (see Table
\ref{table1} and Fig. \ref{fig:neutrons}), further analysis was
estimated necessary to unambiguously characterize the Cu/V
substitution.

Using the VASP package \cite{vasp}, Density functional theory
(DFT) methods applied to electronic structure calculation, based
on the local-density approximation (LDA) and Projector Augmented
Wave (PAW) method, were used for this purpose. The DFT
calculations were performed with periodic boundary conditions on
one unit cell containing 130 atoms and including 2 clusters
located in the two (Cu/V)O$_{3/3}$ layers at $z$=1/4 and $z$=3/4.

\begin{table*}
\caption{Refined structural parameters from Rietveld analysis of
neutron powder diffraction data of \lacuvo at 3 K. B$_{iso}$ is
the isotropic displacement parameter in \AA$^2$. The agreement
factors are 3.38 for the Bragg R-factor and 2.27 for the Rf-Bragg
factor.}
 \label{table1}
 \begin{tabular}{cccccccc}
 \hline\hline
 Atom   &  Wyckoff &  $x$    &  $y$      &  $z$      & B$_{iso}$&  occupancy \\
 \colrule
 La(1)  &  2b   &  0         & 0         &  0        & 0.59(16) &  0.16667\\
 La(2)  &  12i  &  0.2997(2) & 0.0679(3) &-0.0007(3) & 0.56(4)  &  1.00000\\
 La(3)  &  12i  &  0.6148(3) & 0.1480(2) & 0.0124(2) & 0.29(4)  &  1.00000\\
 V(1)   &  2d   &  2/3       & 1/3       & 1/4       & 0.8      &  0.16667\\
 V(2)   &  6h   &  1.0208(5) & 0.1604(5) & 1/4       & 1.97(11) &  0.045(5)\\
 Cu(2)  &  6h   &  1.0208(5) & 0.1604(5) & 1/4       & 1.97(11) &  0.455(5)\\
 Cu(3)  &  6h   &  0.3523(3) & 0.2301(3) & 1/4       & 0.51(5)  &  0.50000\\
 Cu(4)  &  6h   &  0.6092(5) & 0.5841(4) & 1/4       & 1.65(8)  &  0.50000\\
 V(5)   &  6h   &  0.158(8)  & 0.700(7)  & 1/4       & 0.8      &  0.50000\\
 O(a)   &  12i  &  0.4353(4) & 0.0328(4) & 0.0743(4) & 0.62(6)  &  1.00000\\
 O(b)   &  12i  &  0.7524(4) & 0.1164(4) & 0.0766(4) & 0.61(7)  &  1.00000\\
 O(c)   &  12i  &  0.0515(4) & 0.1798(4) & 0.0830(4) & 0.82(7)  &  1.00000\\
 O(d)   &  4f   &  2/3       & 1/3       & 0.0768(8) & 0.48(14) &  0.33333\\
 O(e)   &  12i  &  0.2737(4) & 0.4934(4) & 0.1103(3) & 0.48(5)  &  1.00000\\
 O(1)   &  2a   &  0         & 0         & 1/4       & 2.65(31) &  0.16667\\
 O(2)   &  6h   &  0.3022(6) & 0.0722(6) & 1/4       & 1.37(9)  &  0.50000\\
 O(3)   &  6h   &  0.5657(6) & 0.3726(6) & 1/4       & 0.95(10) &  0.50000\\
 O(4)   &  6h   &  0.2884(5) & 0.3418(6) & 1/4       & 0.64(9)  &  0.50000\\
 O(5)   &  6h   &  0.4519(6) & 0.5571(5) & 1/4       & 0.96(10) &  0.5000\\
 \hline\hline
 \end{tabular}
\end{table*}

\begin{figure}
\includegraphics[scale=0.4]{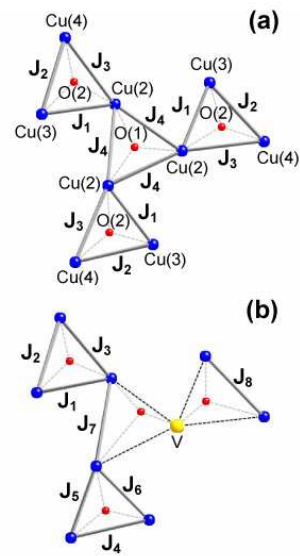}
\caption{(Color online) DFT calculated positions of the Cu (blue
circles), V (large yellow circles), and O (small red circles)
atoms in the 9-spin clusters (a) and 8-spin clusters (b). The
labeling of the different superexchange interactions is used in
section \ref{sec:model}.} \label{fig:substitutions}
\end{figure}

The non substituted (and hence non stoichiometric) structure was
first optimized, and found very close to the one determined from
powder neutron diffraction. Different calculations were then
performed replacing, in the optimized structure, one Cu by a V,
alternatively on each Cu(2), Cu(3) and Cu(4) site of the cluster.
The large energy difference between these calculated structures
first confirmed the localization of the substitution on the Cu(2)
sites : $E_{\mathrm{Cu}(3)}\simeq E_{\mathrm{Cu}(4)} >
E_{\mathrm{Cu}(2)}$ by $\sim 2$ eV, with $E_{\mathrm{Cu}(i)}$ the
energy of the structure with the Cu/V substitution on the Cu$(i)$
site.

In addition, the calculated structure reveals large deformations
around the V substitute. The two first neighbor O of the V get
closer to it and further away from the neighboring Cu (cf. Fig.
\ref{fig:substitutions}). A consequence of this is the apparent
greater isotropic displacement parameter B$_{iso}$ of the
neighboring oxygen on site O(1) in the Rietveld analysis of the
neutron diffractogramms (see Table \ref{table1}).

Calculations with two V/Cu substitutions on the same cluster were
finally carried out and compared to those with only one
substitution. The obtained energy difference between these two
structures, of the order of 1 eV, is in favor of only 1 Cu/V
substitution per cluster. The resulting reasonable assumption of a
negligible number of 7 (or less) Cu clusters therefore leads to a
microscopic population of 2/3 of 9-Cu clusters and 1/3 of 8-Cu/1-V
clusters.

\subsection{\label{sec:Magn}Magnetic properties}

\subsubsection{\label{sec:collspin} Magnetic susceptibility}

\begin{figure}
\includegraphics[scale=0.32]{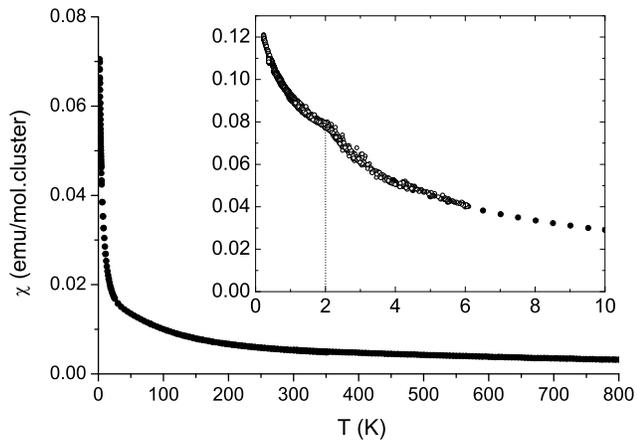}
\caption{La$_3$Cu$_2$VO$_9$ magnetic susceptibility from M/H
measurements: in the high temperature range at 6 T using the BS
magnetometer and at 0.1 T using the Quantum Design MPMS
magnetometer; in the low temperature range using the purpose built
SQUID magnetometer in 0.01, 0.1, 0.2 T (open circles), in good
agreement with the higher temperatures measurements (plain
circles). A discontinuity around 2 K is pointed out by the
vertical line. }\label{fig:chi}
\end{figure}

Magnetization measurements were performed at the Institut Neel, on
two purpose built magnetometers (BS) using the axial extraction
method in the temperature range from 1.6 K to 800 K and magnetic
field up to 10 T and on a more sensitive commercial Quantum Design
SQUID magnetometer (MPMS) in the temperature range 2-350 K and
magnetic field up to 5 T. Additional measurements at lower
temperatures, from 0.23 K to 4 K, using a dilution inserted device
were obtained on another purpose built highly sensitive SQUID
magnetometer.

The thermal variation of the initial magnetic susceptibility
$\chi$ is displayed in Fig. \ref{fig:chi}. $\chi$ was deduced from
the $M~vs.~H$ measurements in small magnetic fields such that the
isothermal magnetization M varies linearly with the magnetic field
H. An anomaly is observed around 2 K. The next section is devoted
to the analysis of the susceptibility in the paramagnetic regime
above the anomaly, the nature of the phase below it will be the
subject of another publication \cite{avenir}.

\begin{figure}
\includegraphics[scale=0.4]{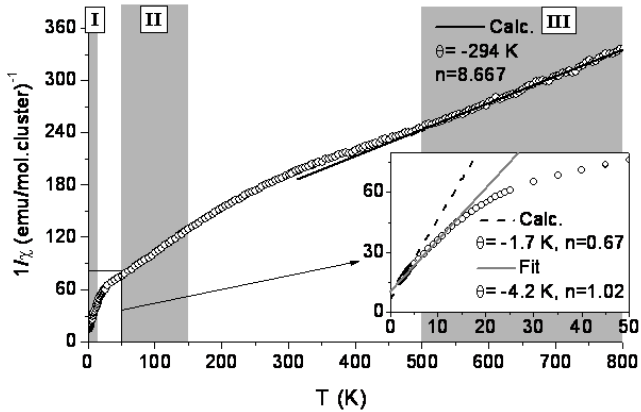}
\caption{La$_3$Cu$_2$VO$_9$ inverse linear magnetic susceptibility
measured in 0.1 T and 6 T below and above 350 K respectively (open
circles). Three regions of distinct magnetic behaviours are
materialized in grey. A low temperature zoom is shown in the
inset. The straight lines in regions I and III are calculated in a
Curie-Weiss model (see text).}\label{fig:invchi}
\end{figure}

The inverse $\chi^{-1}$ of the initial magnetic susceptibility,
shown in Fig. \ref{fig:invchi}, has a peculiar shape, where three
distinct regions, with different (almost) linear slopes at low
(I), intermediate (II), and high (III) temperatures, can be
distinguished. In a first analysis following ref. \cite{vander},
regions (I) and (III) were assumed to correspond to two distinct
paramagnetic regimes of interacting magnetic entities accounted
for, in the mean field approximation, by a Curie-Weiss law
$\chi=C/(T-\theta)$ with the Curie-Weiss temperature $\theta$. The
Curie constant, $C=ng^2\mu_BS(S+1)/3k_B$ with the Land\'e factor
$g$=2, is evaluated for 1/3 of 8-spin cluster + 2/3 of 9-spin
clusters, i.e. 8.667 Cu$^{2+}$; $n$ is therefore the number of
$S$=1/2 spins per cluster. The resulting Curie-Weiss analysis are
shown in Fig. \ref{fig:invchi} and the numerical values are
reported in Table \ref{tab:CW}: $n$, $\theta$, the effective
moment $\mu_{eff}=g\sqrt{S(S+1)}$, and the temperature range of
the best fits corresponding to the linear portion of regions (I)
and (III). The slope of the high temperature region (III) is well
described above 500 K with a Curie-Weiss fit using the expected
value $n$ of 8.667 $S$=1/2 spins per cluster in the paramagnetic
regime (Fig. \ref{fig:invchi}). The resulting large negative
Curie-Weiss temperature for this regime, $\theta=-294$ K, denotes
strong intra-cluster antiferromagnetic couplings between these
spins.

In the low temperature region (I), it is more difficult to isolate
a linear regime in $1/\chi$, which rather presents a continuous
curvature (Fig. \ref{fig:chi}). A forced Curie-Weiss fit in a
reduced temperature range yields approximately one $S$=1/2 spin
per cluster. The reduction of the Curie constant, by a factor
close to 9, suggests that the magnetic entities in the low
temperature range are collective $\tilde S$=1/2 pseudo-spins,
resulting from the entanglement of the wave functions of the
paramagnetic spins within each cluster at high temperature. These
pseudo-spins seem to be weakly antiferromagnetically coupled to
each other since $\theta=-4.2$ K in this region, which is about
two orders of magnitude weaker than the $\theta$ value of region
(III) corresponding to the intra-cluster coupling. This value,
close to the one at which the susceptibility presents an anomaly,
suggests that this last feature is related to the inter-cluster
coupling which becomes effective at this temperature (cf. section
\ref{sec:collspin}). We will come back to the analysis of the low
temperature regime in section \ref{sec:theory} using a more
detailed description of the clusters.

\begin{figure}
\includegraphics[scale=0.45,bb=0 410 562 800]{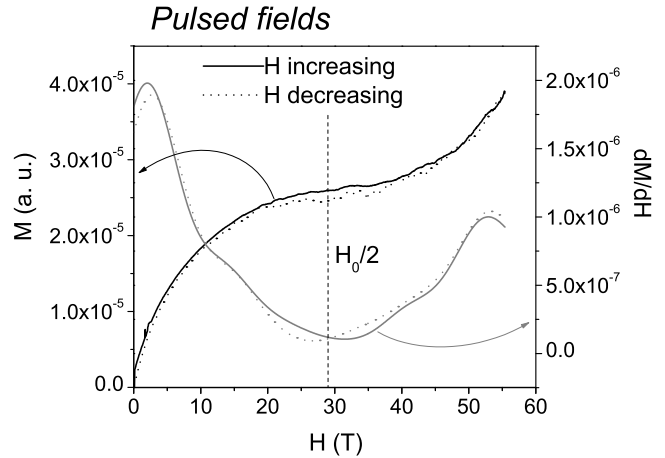}
\caption{Magnetization isotherm measured around T=1.4 K in
increasing and decreasing pulsed field up to 55 T, and its field
derivative. The dotted line indicates the middle of the
magnetization plateau, which is related to the Zeeman crossing of
the two first levels of the system with $\Delta$S=+1 (see inset of
Fig. \ref{fig:HFextr}).}. \label{fig:HFpulsed}
\end{figure}

\begin{figure}
\includegraphics[scale=0.45,bb=0 390 562 800]{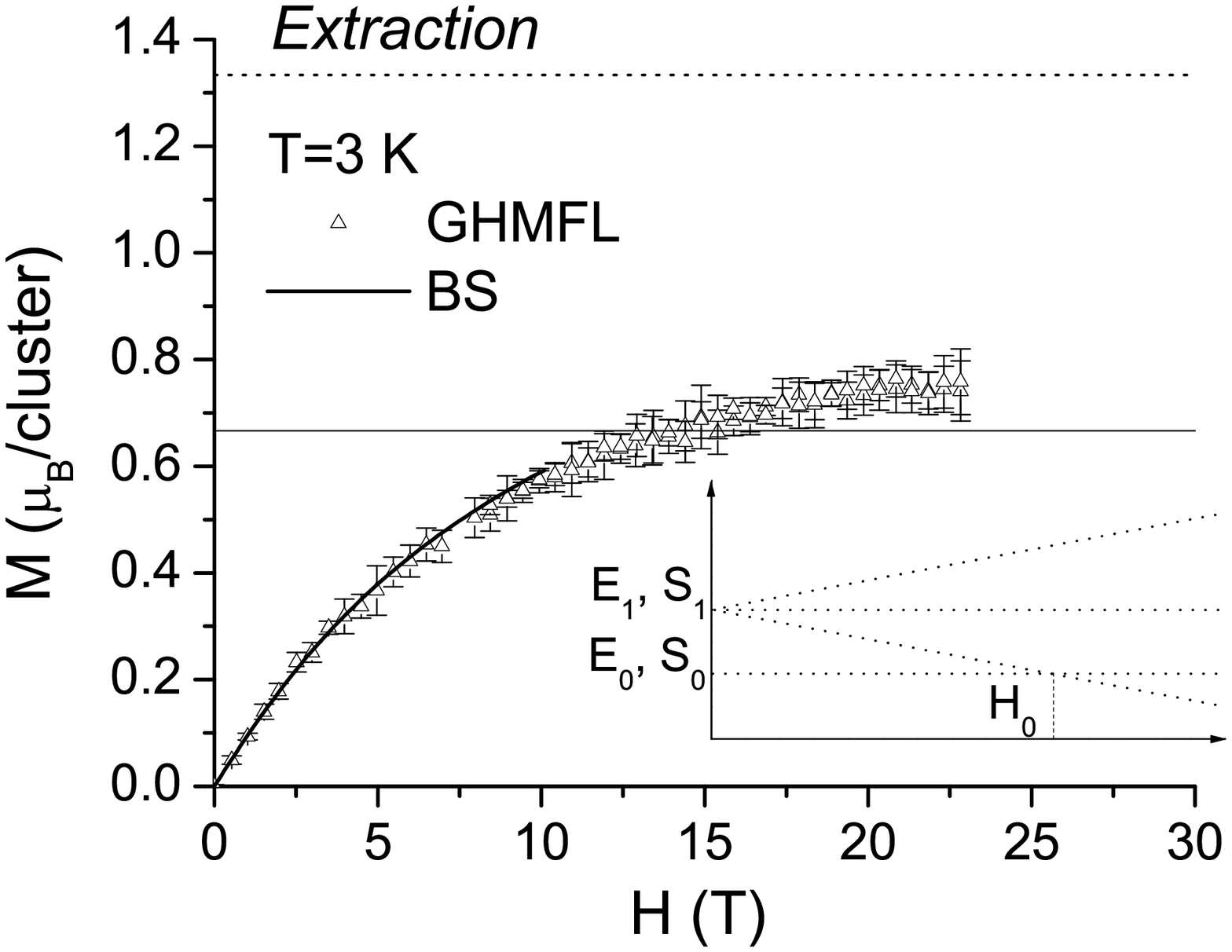}
\caption{High field magnetization isotherms in meaningful units,
measured by extraction at 3 K , and compared to the BS
magnetization measurements. The solid and dotted horizontal lines
point out the calculated saturation value of the ground state and
of the first excited state magnetizations (see section
\ref{sec:theory}). The Zeeman diagram corresponding to the
field-induced crossing of these two levels is shown in the
inset.}. \label{fig:HFextr}
\end{figure}

Concerning the intermediate temperature region (II), although
$1/\chi$ could also be believed to vary linearly from 40 to 200 K
\cite{vander}, a Curie-Weiss analysis does not have any physical
meaning since no collective magnetic entities are formed, as will
be shown in the theoretical analysis of section \ref{sec:theory}.

\begin{table}
\begin{tabular}{ccccc}
\hline Region & T range (K) & $\theta$ (K) & $\mu_{eff}$ $(\mu_B)$
& $n$ \\ \hline
I & 2-11 & $-4.2 \pm 1.1$ & 1.732 & $1.02 \pm 0.19$ \\
III & 500-850 & $-294 \pm 73$ & 1.732 & 8.667
\\ \hline
\end{tabular}
\caption{Curie-Weiss parameters of the best fits accounting for
the inverse susceptibility in the two quasi-linear regimes at low
and high temperature. In the high temperature region, the number
of spins $S$=1/2 per cluster was hold constant and equal to its
expected value of 8.667.} \label{tab:CW}
\end{table}

\subsubsection{\label{sec:highfield}High field magnetization}

High field magnetization measurements were performed using pulsed
fields at the Laboratoire National des Champs Magn\'etiques
Puls\'es (Toulouse). The pulsed field, which can reach 55 T, is
produced by the discharge of a capacitor bank into a resistive
copper coil. The signal in the presence of the sample is
proportional to the time derivative of its magnetization. The
$M~vs~H$ curve obtained at T $\approx$ 1.4 K is displayed in Fig.
\ref{fig:HFpulsed}. It shows a saturation plateau centered on the
inflexion point $H_0/2\approx$ 29 T, indicated by the minimum in
the magnetization field derivative. Note that the influence of the
weak inter-cluster interactions should be irrelevant at such high
magnetic fields. An upturn of the magnetization occurs at higher
magnetic fields, suggesting that the plateau should end at about
$H_0$ = 58 T. This is attributed to a Zeeman crossing of the first
excited magnetic energy levels with the ground level of the
cluster system (cf. inset of Fig. \ref{fig:HFextr}). The
quantitative analysis of this crossing will be further discussed
in section \ref{sec:theory}.

Since the pulsed field technique does not allow absolute
measurements with a good accuracy, complementary measurements at 3
K were performed at the Grenoble High Magnetic Field Laboratory
(GHMFL) on a magnetometer using the axial extraction under a
magnetic field up to 23 T produced by a 10 MW resistive magnet. An
absolute calibration was performed using a Ni sample. The
resulting curves presented in Fig. \ref{fig:HFextr}, in agreement
with the BS lower fields measurements, show a tendency to saturate
at a value close to 0.67 $\mu_B$/cluster, the expected value for
antiferromagnetic 8- and 9-spin clusters in the (1/3)/(2/3) ratio
(cf. section \ref{sec:theory}).

\subsubsection{\label{sec:SpecHeat}Specific heat}

\begin{figure}
\includegraphics[scale=0.32]{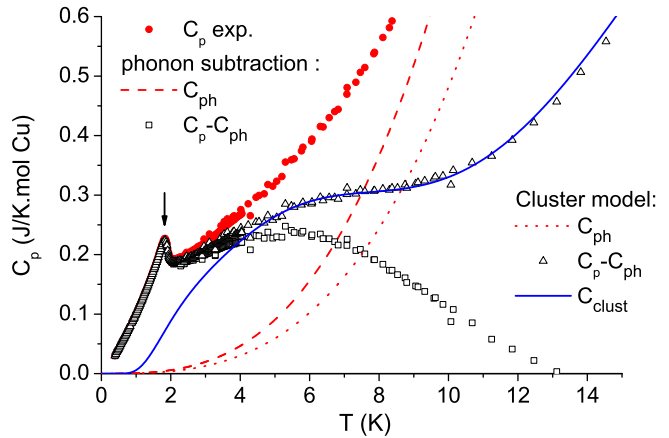}
\caption{(Color online) Specific heat measurements at low
temperature (circles) analyzed with two models: (i) The phonon
contribution (dashed line) is obtained from a fit of $C_p/T$ as a
function of $T^2$ using the low temperature approximation
$C_p=\beta_1 T^3+\beta_2 T^5$ in the range [12-28 K]. The
resulting magnetic signal is represented with square symbols. (ii)
The second model is based on the multi-$J$ spin cluster
description reported in table \ref{tab:multiJ} of section
\ref{sec:theory}. The lattice contribution of the form
$C_p=\beta_1 T^3+\beta_2 T^5$ (dotted lines) is adjusted such that
the remaining signal (triangles) coincides with the calculated
magnetic specific heat (solid line) above 5 K.} \label{fig:specht}
\end{figure}

Specific heat $C_p$ in zero magnetic field was measured at the
CEA-Grenoble on a commercial calorimeter (Quantum Design PPMS)
using the pulsed relaxation method and equipped with an $^3$He
refrigerator allowing to reach 400 mK. Its low temperature
behavior is shown in Fig. \ref{fig:specht}. An anomaly is clearly
visible at 1.82 K, coinciding with the one observed in the initial
magnetic susceptibility $\chi$. In the absence of a non-magnetic
isostructural compound, a determination of the magnetic part of
the specific heat in the full temperature range is difficult.
However, the lattice contribution is often evaluated by a fit of
$C_p/T$ versus $T^2$ in a temperature range where it is assumed
that the magnetic signal is negligible and that the following low
temperature approximation for the phonon contribution is valid:
$C_p=\beta_1 T^3+\beta_2 T^5$ with $\beta_1={12\pi^4 \over 5}Nk_B
/\Theta^3_D$ where $\Theta_D$ is the Debye temperature and N is
the number of vibrating ions. In the present case, the best fit in
the 12-28 K temperature range yields the values:
$\beta_1=(7.3\pm0.09) 10^{-4}$, as from which we get $\Theta_D
\approx 271~K$, and $\beta_2=(-2.3\pm 0.1) 10^{-7}$. The assumed
magnetic part of the specific heat, remaining after subtraction of
this contribution, presents a broad signal centered around 6 K, as
shown in Fig. \ref{fig:specht}. This signal is interpreted as the
signature of a Schottky anomaly due to the presence of discrete
low energy levels associated to the spin clusters. This analysis
will be deepened using the finer description of the pseudo-spins
obtained from the calculation discussed in section
\ref{sec:theory}.

\subsubsection{\label{sec:Neutron}Inelastic neutron scattering}

In order to probe the energy spectrum of the assembly of
collective spins in \lacuvo, inelastic neutron scattering
measurements were performed on the IN4 time-of-flight spectrometer
at the Institut-Laue-Langevin with an incident neutron beam of
energy 16.9 meV (i.e. of wavelength $\lambda$=2.2 \AA), and an
energy resolution of 0.77 meV. The spectra were corrected from the
background contribution and the detector efficiency was calibrated
using a vanadium sample.

\begin{figure}
\includegraphics[scale=0.4,bb=0 140 572 780 ]{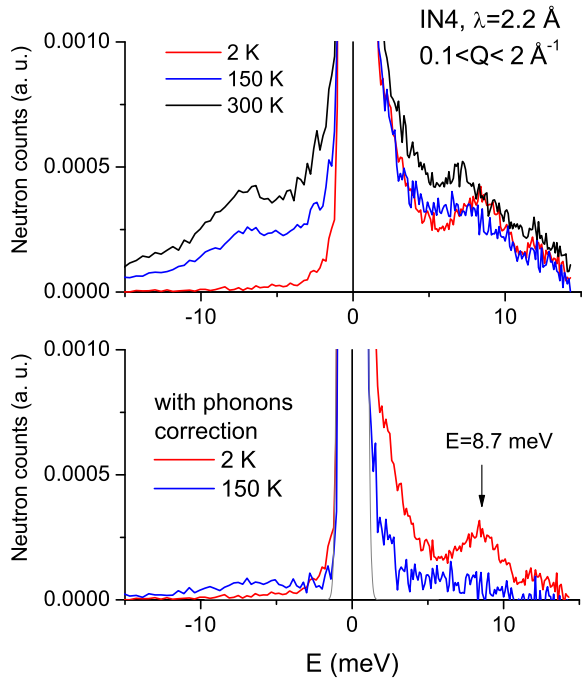}
\caption{(Color online) Inelastic neutron scattering measurements.
Upper frame: Energy spectra at 2, 150 and 300 K, integrated on the
$Q$ interval [0.1, 2] \AA$^{-1}$. Lower frame: 2 and 150 K spectra
corrected from the phonon contribution using the 300 K spectrum
(see text).} \label{fig:neutron}
\end{figure}

The energy response recorded at 2, 150 and 300 K integrated over
the small momentum transfer $Q$, more precisely in the range [0.1,
2] \AA$^{-1}$, where the magnetic scattering is expected to be the
largest are displayed in Fig. \ref{fig:neutron}. As apparent from
the increase of the signal with increasing temperature and with
increasing $Q$, there is an important contribution from the
scattering by the phonons in the investigated energy range (up to
15 meV on the neutron loss energy side). A way of removing it is
to substract the spectrum at high temperature T$_H$=300 K
renormalized by the appropriate thermal factor
$(1-\exp(-{\hbar\omega\over k_B{\rm
T}_L}))/(1-\exp(-{\hbar\omega\over k_B{\rm T}_H}))$ from the
spectra at lower temperatures T$_L$. This treatment might
underestimate the magnetic contribution since magnetic signal
arising from transitions between energy levels can still be
present at high temperature, especially if the energy spectrum of
the system is dense and extended in energy. This analysis allowed
nevertheless to visualize in the corrected spectra a broad bump
around 8.7 meV at 2 K , which largely decreases at 150 K. The
magnetic origin of this feature was confirmed by its slow decrease
with increasing $Q$ from 1 to 4 \AA$^{-1}$, once corrected from
the phonons. An additional signal at 2 K seems to be present in
the foot of the elastic peak. The nature of this signal is hard to
establish from this sole measurement since it could as well be
quasielastic or inelastic with an energy gap of the order of the
energy resolution.

\section{\label{sec:theory}Analysis and discussion}

The crystallographic study established that each \lacuvo plane can
be described as a triangular lattice of randomly distributed 1/3
of 8-spin and 2/3 of 9-spin clusters with antiferromagnetically
interacting localized $S$=1/2 spins on each Cu. The magnetic
properties of \lacuvo\ in the range of temperature $T \gtrsim$ 2 K
should be dominated by a physics of isolated clusters, since the
inter-cluster interactions are two orders of magnitude smaller
than the intra-cluster ones. In the following, we shall probe this
regime through calculations by exact diagonalization of 8-spin and
9-spin clusters model. The intra-cluster interactions
parameterizing these calculations will be adjusted from comparison
with the experimental results. These impose the following
constraints: first, the calculation should reproduce the measured
thermal variation of the initial magnetic susceptibility and the
magnetization processes in high magnetic field, including the
saturation plateau and the Zeeman crossing at the largest
available magnetic field. Second, the calculated energy spectrum
of the system should account for the energy excitations revealed
by inelastic neutron scattering and the magnetic specific heat.
The nature of the resulting pseudo-spin state at low temperature
will then be discussed.

\subsection{\label{sec:model} Cluster calculation versus experiments}

\begin{figure}
\includegraphics[scale=0.35,bb=20 30 814 572]{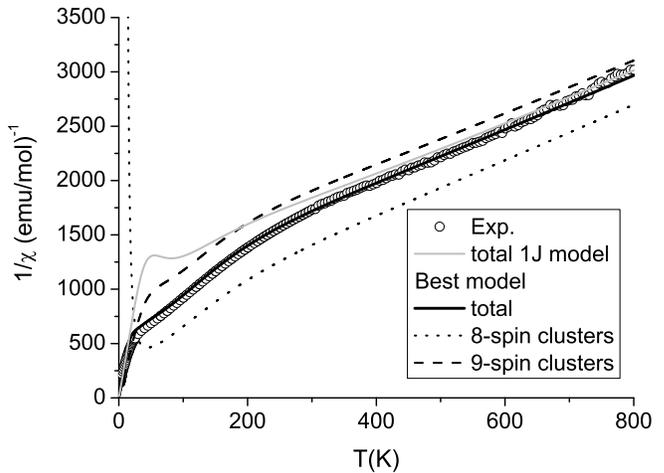}
\caption{Comparison of the inverse of the measured susceptibility
(circles) with calculated ones (line) from a single $J$ model
(grey) and from the multi-$J$ model of table \ref{tab:multiJ}
taking into account 8-spin (dotted) and 9-spin clusters (dashed).}
\label{fig:fitchi}
\end{figure}

The calculation of the \lacuvo magnetic properties are obtained
through exact diagonalisation of the Hamiltonian of the model
system where each 8-spin and 9-spin cluster is described by the
Heisenberg Hamiltonien
\begin{eqnarray}
\label{eq:heis} \mathcal{H}=- \sum_{ \langle ij \rangle} J_{ij}
\mathbf{S}_i \cdot \mathbf{S}_j
\end{eqnarray}
\noindent where the $J_{ij} < 0$, labeled in Fig.
~\ref{fig:substitutions}, stand for all nearest neighbor
antiferromagnetic interactions between the spins $i$ and $j$ of
the cluster (Fig.~\ref{fig:substitutions}). In a single-$J$ model
assuming that all $J_{ij}=J=-445$ K, the ground states of the 9-
and 8-spin clusters are found to be associated with a total spin
$\tilde S$=1/2 and $\tilde S$=0, respectively. This yields in
average 0.667 $S$=1/2 spin per cluster at low temperature, which
is a bit smaller than the value extracted from the Curie-Weiss fit
of the paramagnetic regime at low temperature (dotted line in the
inset of figure \ref{fig:invchi}). The magnetic susceptibility and
other thermodynamics properties of the system were computed and
compared to the experimental data. A single-$J$ model does not
account for the initial magnetic susceptibility as shown in Fig.
\ref{fig:fitchi} where one can see strong deviations from the
experimental data below T $\approx 300$ K down to low
temperatures. This demonstrates the need to consider different $J$
values, that will be expressed, in the following, in units of
$J=-445$ K.

\begin{figure}
\includegraphics[scale=0.32,bb=20 30 814 572]{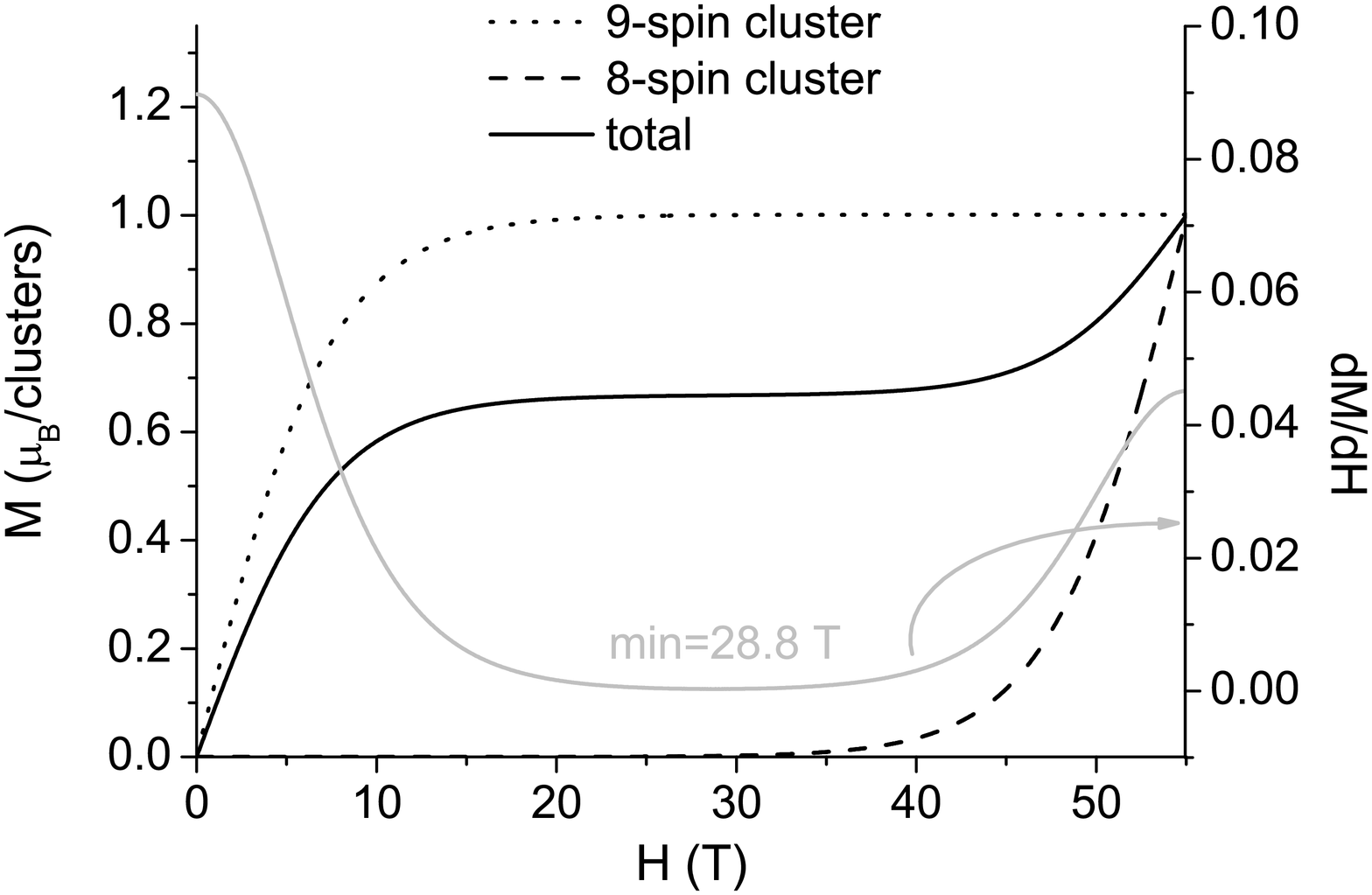}
\caption{Calculated total M(H) in the multi-$J$ model of table
\ref{tab:multiJ}, with the 8-spin (dotted) and 9-spin (dashed)
contributions, at 5 K. The grey line is the field derivative of
the total M(H).} \label{fig:MHcalc}
\end{figure}

To push further the analysis, some additional assumptions are
required in order to reduce the large number of non-equivalent
first-neighbor interactions : 4 and 8 distinct $J_{ij}$ values for
the 9- and 8-spin clusters respectively, as shown in Fig.
\ref{fig:substitutions}. In the 8-spin clusters, the main
consequence of the distortion induced by the Cu/V substitution is
an elongation of the $J_7$ and $J_8$ exchange paths (see section
\ref{sec:substitutions} and Fig. \ref{fig:substitutions}). These
exchange parameters are thus expected to be much weaker than all
the others in the 8- and 9-spin clusters. We therefore assume a
3-J model for the 8-spin clusters, i.e a model defined within the
($J_1 = \ldots = J_6, J_7, J_8$) parameter space. Concerning the
9-spin clusters, there are no straightforward assumptions that
could be made and the 4-J model described in
Fig.~\ref{fig:substitutions} is adopted.

The measured magnetization processes under magnetic fields provide
with meaningful constraints over the $J_7$ and $J_8$ parameters.
The saturation of the magnetization, corresponding to the
alignment of the $\tilde S$=1/2 pseudo-spin of the 9-spin clusters
anti parallel to the applied magnetic field, is reproduced already
within the single J model. Assuming weaker $J_7$ and $J_8$, the
first excited states which will be field-driven to energy lower
than that of the zero field ground state are those belonging to
the 8-spin cluster excited manifold of states with total spin
$\tilde S$=1 and describing antiparallel alignment to the applied
magnetic field. The corresponding Zeeman crossing experimentally
inferred to occur at about 58 T then imposes that $J_7$ and $J_8$
cannot take values smaller than 0.28 and 0.16, respectively. In
order to get the inflexion point $H_0/2$ around 29 T, one of these
parameters must at least take the above limit values, as shown in
Fig. \ref{fig:MHcalc} for $J_7$=0.28 and $J_8$=0.24. Note that
M(H) has been calculated for T=5 K in order to round the sharp
edges of the calculated magnetization saturation plateau and thus
better reproduce the measurements. This shape difference together
with the fact that the experimental saturation is slightly higher
than the calculated one could be due to a distribution of $J$
values not taken into account in our simplest model (some kind of
disorder, influence of inter-cluster coupling, presence of a tiny
amount of clusters with different Cu/V substitutions etc.)

\begin{figure}
\includegraphics[scale=0.42,bb=10 400 572 814]{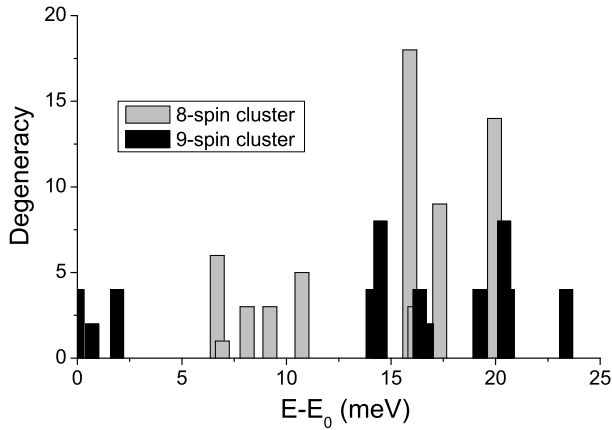}
\caption{8-spin and 9-spin clusters degeneracies as a function of
the energy difference with respect to the fundamental level
(E$_0$) in the low energy region, for the multi-$J$ model of table
\ref{tab:multiJ}.} \label{fig:energy}
\end{figure}

The excitation spectrum associated with this 8-spin cluster model
is shown in figure \ref{fig:energy}. A first bunch of levels is
observed around 8.5 meV above the ground state, and the next bunch
lies at energies greater than 15 meV. At 2 K, this first bunch of
levels will therefore yield an excitation at an energy value in
agreement with the one observed in inelastic neutron scattering
experiment (cf. Fig. \ref{fig:neutron}).

\begin{table}
\begin{tabular}{cccc|ccccccccc}
\hline \multicolumn{4}{c}{9-spin clusters} &
\multicolumn{8}{c}{8-spin clusters}
\\ \hline
$J_1$ & $J_2$ & $J_3$ & $J_4$ & $J_1$ &$J_2$ &$J_3$ &$J_4$ &
$J_5$&$J_6$ &$J_7$ &$J_8$
\\
1.0 & 1.01 & 1.045 & 1.02 & 1 & 1 & 1 & 1 & 1 & 1 & 0.28 & 0.24
\\ \hline
\end{tabular}
\caption{Example of multi-$J$ model that accounts well for all the
experimental data. The 8-spin and 9-spin clusters nearest-neighbor
interactions $J$, labeled as in figure \ref{fig:substitutions},
are given in units of 445 K.} \label{tab:multiJ}
\end{table}

Small variation of the $J$s on the other bonds of the 8-spin and
9-spin clusters are expected from the slight differences in their
Cu-O distances and Cu-O-Cu angles and by the different
coordination environments of the Cu(2) site on one hand and of the
Cu(3) and Cu(4) sites on the other hand (section
\ref{sec:synthesis}). An experimental constraint that gives some
information about this distribution of $J$ values is the broad
bump around 6 K observed in the magnetic specific heat. No
magnetic contribution to the specific heat is observed below 10 K
in the single-$J$ model. Conversely, small differences, no larger
than 5 \%, between the three bonds of the external triangles in
the 9-spin clusters allow the rise of a bump in the calculated
specific heat around 6 K. Note that the low energy spectrum as
well as the shape of the magnetic susceptibility are much less
sensitive to the value of $J_4$ (the inner triangle exchange). In
addition, similar variations of the $J_1$ to $J_6$ bonds in the
8-spin clusters does not yield any feature in the specific heat in
this low temperature range. On this basis, a multi-$J$ model with
$J_1$=1.0, $J_2$=1.01, $J_3$=1.045, and $J_4$=1.02 in the 9-spin
clusters (cf. Table \ref{tab:multiJ}) inferred from the specific
heat, was proposed and found to also well reproduce the inverse
magnetic susceptibility, as shown in figure \ref{fig:fitchi}.

The resulting specific heat is shown in figure \ref{fig:specht}.
In addition to the requested bump around 6 K, the calculation also
reveals the presence of significant magnetic signal at high
temperatures. This implies that the evaluation of the phonon
contribution that was made section \ref{sec:SpecHeat}, assuming
that there is a region where it follows the low temperature Debye
approximation and where the magnetic contribution is negligible,
is incorrect. However, a phonon contribution of the form
$C_p=\beta_1 T^3+\beta_2 T^5$ with $\beta_1=0.00047$,
corresponding to $\Theta_D \approx 314~K$, and $\beta_2$=1.1184
$10^{-7}$ can be adjusted such that the remaining signal coincides
with the calculated magnetic specific heat above 5 K (cf. figure
\ref{fig:specht}).

The low lying energy levels obtained through such small variations
of the exchange parameters on the external triangle in the 9-spin
clusters are shown via the histograms of figure \ref{fig:energy}.
They would also be in agreement with the presence, in the neutron
scattering experiment, of a magnetic inelastic signal at low
temperature in the foot of the elastic peak (cf. figure
\ref{fig:neutron}).

Although not necessary for the matching of the calculations with
the experimental data, small variations of the $J_1$ to $J_6$
exchange parameters in the 8-spin clusters are also expected to be
present in the real system. The exact distribution of all the
8-spin and 9-spin clusters exchange parameters can however not be
accessed unambiguously from the present data. Because of the large
number of adjustable parameters, the multi-$J$ model presented
table \ref{tab:multiJ} is thus not the only relevant, but the main
features of this model, arising from the experimental constraints,
are quite robust : smaller values of $J_7$ and $J_8$ in the 8-spin
clusters with respect to all other first-neighbors parameters, and
distribution of exchange parameter values of the order of less
than 5 \% in the 9-spin clusters. Note that this model gives a
satisfactory agreement with all the experimental data. The inverse
magnetic susceptibility in particular is much better reproduced
than in the single-$J$ model, except for tiny differences at low
temperatures. These last discrepancies could be due to the onset
of inter-cluster correlations at low T that were not included in
the model.

\begin{figure}
\includegraphics[scale=0.32,bb=50 40 814 572]{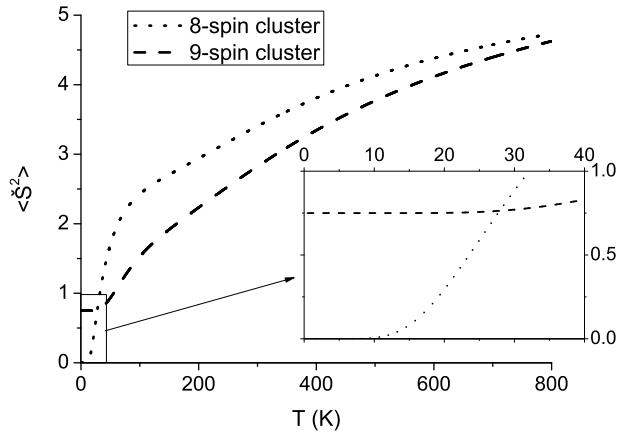}
\caption{\label{fig:S2} Calculated total $\langle \tilde S^2
\rangle$ as a function of temperature for the 8-spin cluster
(dotted lines) and 9-spin cluster (dashed lines) in the multi-$J$
model of table \ref{tab:multiJ}. Zoom of the low temperature range
in the inset.}
\end{figure}

The evolution of these pseudo-spins when lowering temperature can
be probed through the square of the total spin of the cluster
$\langle \tilde S^2\rangle$. Commuting with the cluster
Hamiltonian, $\langle \tilde S^2\rangle$ is a good quantum number
that allows to track the spin value of the entangled state in the
whole investigated temperature range. In the multi-$J$ model, the
pseudo-spins are stabilized below $\approx$ 18 K for the 9-spin
clusters and below $\approx$ 8 K for the 8-spin clusters as shown
by the constant $\langle \tilde S^2\rangle$ values, 3/4 and 0
respectively, observed below these temperatures (cf. inset of Fig.
\ref{fig:S2}). Concerning behavior at higher temperatures, a $1/T$
expansion of the calculation in the region III, confirms that the
system can be described by a Curie-Weiss analysis in the
temperature range $T \gtrsim 550$ K as done in section
\ref{sec:collspin}. In the intermediate temperature region II, no
plateau regime interpretable in terms of entangled spins states
nor a $1/T$ behavior is observed. The variation of the magnetic
susceptibility in this temperature regime (cf. figure
\ref{fig:invchi}) is rather related to the onset of the strong
magnetic intra-cluster correlations and cannot be ascribed to any
building of pseudo spins.

\section{Conclusion}

The \lacuvo oxide compound is shown to be constituted of 8-spin
and 9-spin clusters of $S$=1/2 spins, laid out on 4 vertex-sharing
triangles, building block of the kagom\'e lattice. From
susceptibility, high field magnetization, specific heat and
inelastic neutron scattering measurements analyzed with exact
calculation of spin cluster models, the low temperature
stabilization of collective pseudo-spins $\tilde S$=0 and $\tilde
S$=1/2 in each 8-spin and 9-spin cluster, is revealed. From
Susceptibility and specific heat measurements, the coupling of
these collective entities is evidenced below 2 K. This study opens
the way to original measurements of the individual and cooperative
behaviour of collective quantum entities, for instance a direct
imaging of their wave function using polarized neutron scattering.

\acknowledgments R. Boursier is acknowledged for his help during
the high field magnetization measurements at the GHMFL. The work
at the GHMFL has been supported by the "Transnational Access to
Infrastructures - Specific Support Action" Program - Contract nr.
RITA-CT-2003-505474 of the European Commission.


\end{document}